\documentclass[12pt]{article}
 
\usepackage{amssymb}
\usepackage{graphicx}

\newcommand{\vs}{\vspace{15pt}} \newcommand{\n}{\noindent} 
\newcommand{\rf}[1]{(\ref{#1})}
\newcommand{\ba}{\begin{array}} \newcommand{\ea}{\end{array}}
\newcommand{\be}{\begin{equation}} 
\newcommand{\btb}{\begin{tabular}}\newcommand{\etb}{\end{tabular}}
\newcommand{\ee}[1]{\label{#1}\end{equation}}
\newcommand{\bi}{\bibitem}

\newcommand{\dss}{\displaystyle}
\newcommand{\bfl}{\begin{flushleft}}\newcommand{\efl}{\end{flushleft}}

\newcommand{\al}{\alpha} \newcommand{\bt}{\beta}\newcommand{\g}{\gamma}\newcommand{\G}{\Gamma}  
\newcommand{\de}{\delta}\newcommand{\De}{\Delta}\newcommand{\ep}{\epsilon}
\newcommand{\te}{\theta} \newcommand{\la}{\lambda}\newcommand{\La}{\Lambda}  

\newcommand{\w}{\omega}

\newcommand{\PCU}{{\cal P}_\uparrow}

\newcommand{\Vy}{{\bf V}} 
 
\newcommand{\ey}{{\bf e}}

\newcommand{\ky}{{\bf k}} 
 
\newcommand{\ry}{{\bf r}} 
\newcommand{\uy}{{\bf u}}\newcommand{\vy}{{\bf v}} 
\newcommand{\0}{{\mathbf 0}}

\newcommand{\dg}{\dot{\gamma}}

 \newcommand{\hc}{\hat{c}}

\newcommand{\ra}{\rightarrow} \newcommand{\ol}{\overline}

\newcommand{\tl}{\tilde}

\begin{document} 
\title{On Selleri's ``Weak Relativity''}
\author{Marco Mamone-Capria\\ \small Dipartimento di
Matematica - via Vanvitelli, 1 - 06123 Perugia - Italy \\ \small {\sl E-mail}:
\texttt{mamone@dmi.unipg.it}}\normalsize \maketitle

\small
{\bf Abstract} F. Selleri's main arguments  for the restoration of absolute simultaneity in physics are analysed and shown to be faulty. In particular both the classical Sagnac effect and the recent so-called linear Sagnac effect can be dealt with within special relativity in a natural way. The appeal to the conventionality of simultaneity thesis is also shown to be ineffectual. Other arguments, such as the two spaceships' argument and the block universe argument are briefly examined. Notwithstanding a negative overall assessment, the importance of keeping alive the research on the foundations of relativity is emphasized and Selleri's role in this undertaking appreciated.

\normalsize

\tableofcontents

\section{Introduction}

 Franco Selleri (1936-2013) was an Italian particle physicist who became an expert in the foundations of quantum mechanics first, and then, in the last two decades of his life, a critic of the received view of special relativity (SR). Here is how he described himself at 54, in his first paper devoted to a modified version of the theory of relativity \cite{S95b}:

 \begin{quote}\small
 Franco Selleri graduated in theoretical physics at Bologna University (Italy). He spent several years working at CERN (Geneva), Cornell University (USA). He is presently full professor of theoretical physics at Bari University. His publications include 7 books and over 200 papers. Forty percent of his papers result from collaborations with physicists of 14 countries. He is a member of the scientific boards of five journals, and member of some scientific societies, among which the Foundation Louis de Broglie, the New York Academy of Sciences, and the American Association for the Advancement of Science. He has delivered university courses in seven countries and given invited talks in about sixty international conferences. His interests range from particle physics to the foundations of quantum theory and relativity. Author of the one-pion exchange model in particle physics, he is advocate of a reformulation of quantum physics along the lines dear to Einstein, de Broglie, Bohm and Bell, with a particular emphasis on the validity of local realism in nature. He believes in the objective existence of quantum waves in space and time (together with quantum particles) and has proposed experiments that could lead to their direct detection. He is presently working at a reformulation of the theory of relativity, and the paper published in this issue is the first step in that direction.
\end{quote}\normalsize 

\n
From 1990 to 2012 Selleri published in different venues over 40 contributions on relativity. Clearly he believed that his findings deserved a wide dissemination, and he wanted to promote a new research programme, pleading for a return to absolute simultaneity conceived of as a ``liberation of time from the enslavement to space forced upon it in Minkowski space'' \cite{S04d}.  In the bibliography I append a very representative (though incomplete)\footnote{A more complete list can be found in \cite{rom}, which also provides very rich information on Selleri's professional activity and correspondence with other scientists.} chronological list, with the addition of three books on the foundations of quantum mechanics.  In fact the main arguments are just a few,  and the one Selleri considered as decisive  had to do with rotational motion, particularly as exemplified in Sagnac's experiment. As a brief reference for my treatment one can consult \cite{S10}, which is a final recapitulation of those main arguments; while I did not mean to produce a detailed analysis of all Selleri's relativity papers I listed (and which often have a considerable overlap with each other), I checked the other papers and made occasional references to some of them. 

From Selleri's self-portrait quoted in full above, we see that his research interest in the foundations of the theory of relativity was an outgrowth of his work on the foundations of quantum mechanics. Nonlocality had been established as an essential feature of quantum mechanics by John Bell's inequalities, and this created an obvious difficulty for special relativity, particularly after experimental evidence had been produced indicating that the Bell's inequalities were indeed violated. A considerable part of Selleri's works on this topic tried to fault these experiments with unwarranted reliance on auxiliary assumptions.\footnote{``There are, however, valid reasons that strongly caution against adopting an uncritical belief in the results of these experiments. Indeed [...] the problem of the validity of Bell's inequality still seems open to a variety of solutions.'' (\cite{S90a}, p. 361)} But in case they were confirmed, how to interpret that feature without throwing the baby (relativity) together with the bathwater (locality)? In \cite{mm18} one can find, among other things, an outline of the pre-history of this problem, showing that in fact it had been with quantum mechanics since its birth, well before the Einstein-Podolsky-Rosen paper, and for very good reasons: in its more natural and straightforward presentation quantum mechanics does need absolute synchrony and superluminal influences. This had been pointed out since at least the famous 1927 Solvay conference, but it was (and is) generally neglected in the standard treatments. Heisenberg's solution of the conundrum had relied upon a narrow operationalist view of special relativity: the nonlocal correlations of quantum mechanics cannot be used as signals, and {\sl therefore} - so the argument runs, untiringly and uncritically repeated by countless authors from then on - it cannot contradict the Poincar\'e-Einstein (PE) synchronization, or standard light signal syncronization.\footnote{For justification of this labeling, see section 1 of \cite{mm01}.} 

Here is the plan of this paper. The basic concepts of the operational definition of simultaneity are explained, and Einstein's lack of emphasis on it is pointed out. Sagnac's rotating circular platform experiment is presented, together with its classical treatment. After quoting Selleri's main statements on the supposed inconsistency of special relativity (SR) with the Sagnac effect, two different approaches to the relativistic treatment of the experiment are expounded, one using rotating coordinates, the other focusing on the rotating measuring device. Both approaches lead to the same formula for the time delay, and this formula closely approximates the classical prediction (up to first order in $\bt$). The main objection to these arguments is based on confusing the {\sl instantaneous} speed of light for a rotating observer (in SR this is $c$, of course) and the {\sl average} speeds of the clockwise and counterclockwise photons along the rim of the platform: these average speeds {\sl are} different, one being superluminal while the other is subluminal. The valid part of Selleri's argument exploiting the conventionality of simultaneity thesis in favour of absolute simultaneity is stated as a proposition and proved. The zero acceleration ``discontinuity'' argument is then stated and refuted. The so-called linear Sagnac effect is often used to argue that rotation is not the key to Sagnac effect. However, it is shown that whether standard or non-standard simultaneity is used, the predicted effect remains the same. Selleri's proposal of a ``weak relativity'' is discussed and its inner conflicts (partially acknowledged by him) exposed. A section appreciating Selleri's  role in reviving the debate on the foundations of SR concludes the paper.

\section{Simultaneity, aether and Einstein}

Can the special relativity's concepts of time and causality be reduced to the establishment of a time coordinate by PE-synchronization in an inertial system? In \cite{ei05} Einstein made two relevant points: 1) he stated that the traditional aether had become ``superfluous'', and 2) he called the outcome of the PE-synchronization a {\sl definition}  of the time coordinate in the stationary system. Now a `definition' is in itself a convention. Hans Reichenbach \cite{re57} elaborated this point by surmising that if stationary ideal clocks are located along a straightline, and if we want to synchronize them all from one of them, say $O$, by using light-signals, then we can appeal, as an experimental matter of fact, only to the {\sl two-way} constancy of the speed of light. From this it follows that there exists a continuum of legitimate synchronizations along the given line. More exactly, if at the $O$-time $t_1$ a signal is sent to another clock $O'$ along the line, and is reflected back to $O$ at the $O$-time $t_2$, then the admissible conventions for the time $t'$ of the reflection event that $O'$ should be settled to are:

\[ t'= t_1 + \ep (t_2-t_1),\, \mbox{where}\, \ep \in ]0,1[. \]

\n 
The standard choice is $\ep = 1/2$ for all $O'$, corresponding to the assumption of the costancy of the {\sl one-way} speed of light, and is embodied in the standard Lorentz transformations.. 

More generally, one can prove \cite{mm01}that if we $(\ol{\ry},\ol{t})$ are the position-time coordinates of a coordinate system (cs) $\ol{\phi}$ obtained by the standard synchronization, all other affinely related cs's $\phi$, at rest  with respect to $\ol{\phi}$ and having the same space-time origin, arising from different synchronizations are related to $\ol{\phi}$ by:

\be \ry = \la\ol{\ry}, \; t = \la (\ol{t} + \ky\cdot\ol{\ry}), \mbox{where} \; \la >0, |\ky|< 1/c.  \ee{nons}

Now the set of Minkowski cs's (which is a $\PCU^+$-orbit in the set of space-time coordinatizations, where $\PCU^+$ is the proper orthochronous Poincar\'e group \cite{mm16}) has no unique simultaneity, but one can extract from it a subset of cs's sharing the same simultaneity, and comprising a cs for every admissible velocity $\Vy$ with respect to $\ol{\phi}$ (that is, $|\Vy|<c$). In other words, if we construe SR's concept of time as constrained by the two-way constancy of the speed of light only, then we can adopt a convention on one-way speeds of light restoring absolute simultaneity, identified as one out of the infinitely many Minkowski synchronies.

But why should we? What can justify the decision of singling out {\sl one} Minkowski synchrony? In explaining what had motivated him to abandon special relativity and to create general relativity \cite{ei67}, Einstein had complained that in special relativity the two Newtonian absolutes (absolute time and space) had just fused into a single one (absolute space-time) - not a great bargain, from a `relativistic' (or more exactly Machian) point of view. Obviously he would not have been happy with a return to absolute synchrony and with something very close to a reinstatement of the ``superfluous'' aether. In fact, notwithstanding his familiarity with Reichenbach's work (and person), it never occurred to him, as much as I am aware, to point out during the following half-century that within special relativity there was a legitimate way to recover absolute synchrony. My opinion is that it was clear to him that in order to justify such a return to absolute synchrony a new, previously unknown, physical ingredient should have been added to SRspecial relativity, breaking down the Poincar\'e symmetry.  According to some authors, from Georges Sagnac (1869-1928) to Selleri, this is precisely where the interpretation of rotational motion within SR should lead us. Einstein clearly disagreed.

I shall discuss the main arguments advanced by Selleri, which are still debated in the literature (for instance, see \cite{sp17, sp19}).

\section{Sagnac's experiment}

Let us consider in classical physics a rotating circular platform, in an aether cs,  carrying a clock coupled with an emitter and an interference detector on its rim. We shall use the shorthand `ced' (= clock-emitter-detector) for such a device. Suppose the platform's radius is $r$ and that by a system of mirrors one can have monochromatic light signals travelling along the rim in opposite directions. We want to compute the delay in the return times to the ced of two opposite signals. 

\begin{figure}[htp]
\centering
\includegraphics[totalheight=0.2\textheight]{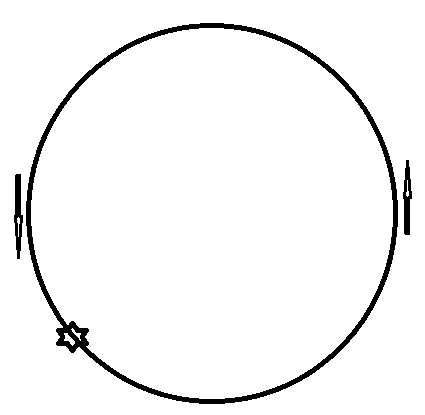}
\caption{Sagnac effect - The star is the clock-emitter-detector (ced)}
\end{figure}

Let the angular speed of the platform  be $\w\, (> 0)$ with respect to the aether cs. To return to the source by circling around the rim the signals will take times $T_ \pm$ satisfying:

\be 2\pi r \pm \w r T_{\pm} = c T_{\pm},\ee{add}

\n
Clearly, in order to have both signals come back to the ced we need assume $\w r < c$, and if we put $\ell := 2\pi r$, $V: = \w r$, $\bt: = V/c$, we obtain:

\[ T_\pm = \frac{\ell}{c} \frac{1}{1 \mp \bt},\]

\n
Thus the {\sl classical} time difference (notice that it does not matter, classically, whether we measure it in the aether cs or in the rotating cs) is:

\be (\De T)_C = T_+ - T_- = \frac{2\ell}{c} \frac{\bt}{1 -\bt^2} = \frac{2\ell}{c}\bt + o(\bt) .  \ee{sag1}

\n 
and the corresponding fringe shift for a monochromatic light of wavelength $\la$ is

\[ \#f = \frac{c \De T}{\la} = 2 \frac{\ell}{\la} \bt + o (\bt). \]

In this sense one can say (as is usually stated  in the literature) that the time delay in Sagnac interferometer is a ``first order effect'' - first order, that is, with respect to the aberration parameter.  However, one must be careful when this kind of classification is used to rate experimental accuracy: given the small rotational velocity in Sagnac's experiment, the final fringe shift was 0.07, to be compared with the predicted fringe shift 0.4 of the original Michelson-Morley experiment; as to the aberration coefficients involved, they were, respectivley

\[ \bt_S = 10^{-7}, \;   \bt_{MM} = 10^{-4}, \]

\n 
so we might say that, as far as the relative effect size is concerned, the Michelson-Morley experiment was a `first-order effect' with respect to Sagnac's, and Sagnac's was a `second order effect' - that is, a more delicate effect. If we assume the `null' outcome of the Michelson-Morely experiment, we can take \rf{sag1} as a good approximation of a Sagnac experiment performed (as the original one) on a `sufficiently) inertial' cs based on the Earth. 

Equation \rf{sag1} can be reformulated in terms of the circle area $A := \pi r^2$:

\be \De T = \frac{4A\w}{c^2} \frac{1}{1 -\bt^2} = \frac{4A\w}{c^2} + o(\bt).  \ee{sag2}

It follows that the angular speed can be derived, to first order, from the fringe shift as:

\[ \w \approx \frac{\la c}{4 A}\# f. \]

In this sense it is correct to conceive of Sagnac's rotating interferometer as an optical Foucault's  pendulum. And as the Foucault's pendulum can be construed as providing indirect evidence for an absolute space, so the Sagnac's experiment can be taken as indirect evidence for an aether. Actually, this is how Sagnac interpreted his experiment, as is clear from his paper's  title: ``The luminous aether proven by the aether's relative wind effect in a uniformly rotating interferometer''.

Notice, however, that the classical Foucault's pendulum proves that the Earth is rotating with respect to a local inertial cs; if we perform the experiment on a platform rotating with respect to a different local inertial cs (with the angular velocity of the Earth's rotation), we would find the same effect. This shows that the Foucault's pendulum cannot single out the subclass of all Galilean cs's which are absolutely at rest. Just as much can be said, of course, of other classical experiments involving inertial forces such as Newton's bucket. 

When we pass to SR we have the additional difficulty that this theory denies that there is any need for postulating an aether.  However, we can say (and on this both Einstein and Weyl agreed) that an avatar of `aether' in SR is the very 4-dimensional metrical structure.\footnote{A detailed historical reconstruction, with extended original quotations, is \cite{ko}.}

\section{How to deal with the Sagnac's effect in SR}

Sagnac's experiment has been considered by Selleri since 1996 as the foremost argument against SR \cite{S96a}. In 2004 he went so far as to write \cite{S04c}:

\begin{quote}\small
{\sl Only in one case (Sagnac effect) can one assert that it is impossible to deduce a formula in agreement with experiments from the TSR} [= SR], while in the other cases it is impossible to give the mathematical symbols a reasonable physical meaning.
\end{quote}\normalsize

\n
Eight years later, in his last paper on the Sagnac's experiment \cite{S12}, Selleri wrote (italics added):

\begin{quote}\small
Most textbooks deduce the Sagnac formula in the laboratory, but say nothing about an observer on the rotating platform. {\sl Special relativity is self-contradictory, as it predicts a null effect on the platform, but a nonzero value if the platform is studied from the laboratory}.
\end{quote}\normalsize

\n
This is a very serious charge. But can it be upheld? To those who, like Selleri, surmise that Sagnac's effect is a stumbling block for SR one can show that there are several ways to deal with the Sagnac's effect in SR, all giving the very same formula. We shall describe two of them, the first one being a refinement of the approach adopted by Paul Langevin in 1921 and 1937 \cite{lan21, lan37}.

\subsection{Rotating coordinates}

Let us fix a Minkowskian cs $\phi$; in order to represent the setting of the experiment, one way is to start with spatially cylindrical coordinates:

\[ x'^1 = r\cos\eta, \; x'^2 = r\sin\eta, \: x'^3= z, \; t'= t, \]

\n
in which the Minkowski metric has the following form:

\be ds^2 = dr^2 + r^2 d\eta^2+ dz^2 - c^2 dt^2 . \ee{min_cyl}

If we want to describe a platform rotating around the $x^3$-axis with angular speed $\w>0$, that is $\eta = \te + \w t$, then we have   

\be ds^2 = dr^2 + r^2 d\te^2+ dz^2 - c^2 (1-\frac{\w^2 r^2 }{c^2})dt^2 + 2\w r^2 d\te dt. \ee{min_rot}

Of course, according to the Principle of Proper Time of SR (\cite{mm01}, \S 11), in Sagnac's experiment the ced on the platform's rim does not show time $t$, but its proper time $\tau$, which satisfies

\[ d\tau = \sqrt{1-\frac{\w^2 r^2}{c^2}} dt. \]
 
\n
Here we assume, like in the classical case but for a more fundamental reason, that the relationship between angular speed and radius is:

\be \w r < c, \ee{bou}

\n
since, if the platform is (as it must be) a material structure, none of its points can move with a speed equal or bigger than $c$. By introducing $d\tau$ into \rf{min_rot}, we get:  

\be ds^2 = dr^2 + r^2 d\te^2+ dz^2 - c^2 d\tau^2  + \frac{2\w r^2 d\te d\tau}{\sqrt{1-\w^2 r^2/c^2}}. \ee{min_rot_1}

In order to compute the angular speed of the circulating light signals with respect to the rotating platform, one has to solve the quadratic equation:

\[ r^2 \left(\frac{d\te}{d\tau}\right)^2 + 2\frac{\w r^2}{\sqrt{1-\w^2 r^2 / c^2}}\frac{d\te}{d\tau} - c^2 = 0, \]

\n
whose solutions are

\be \frac{d\te}{d\tau} = \left\{ \ba{cc} \dss\frac{c/r -\w}{\sqrt{1-\w^2 r^2/c^2}} \\ [8pt]  -\dss\frac{c/r +\w}{\sqrt{1-\w^2 r^2/c^2}} \ea\right. . \ee{angs} 

It follows that the two periods, as measured by the rotating ced, are:

\be T_\pm = 2\pi \frac{\sqrt{1-\w^2 r^2 /c^2}}{c/r \mp \w} = \frac{2\pi r}{c} \frac{\sqrt{1-\w^2 r^2/ c^2}}{1 \mp \w r/c}, \ee{reltim}

\n
and the {\sl relativistic} time difference, as measured by the ced, is:

\be (\De T)_R = \frac{2\ell}{c}\frac{\bt}{\sqrt{1-\bt^2}} = \frac{4 A \w}{c^2}\frac{1}{\sqrt{1-\bt^2}}. \ee{sag_rel}

By comparing \rf{sag1} with \rf{sag_rel} we obtain:

\[ (\De T)_C - (\De T)_R  =
 \frac{4A\w}{c^2}(\frac{1}{1-\bt^2} - \frac{1}{\sqrt {1-\bt^2}}) = \frac{2A\w}{c^2} \bt^2 + o (\bt^2) \]

Thus we can conclude that not only SR deals adequately with the Sagnac setting, but that predicts a result exceedingly close to the classical one, in particular a non-null result as measured by the rotating ced

\subsection{Rotating measuring device}

Some authors object to the use of rotating coordinates since this tastes too much of `general' as opposed to `special' relativity. I think that the root of this objection lies in a misunderstanding of the relationship between the two theories (cf. \S 9), but it is not necessary to enter into this issue, since there is indeed a way of treating Sagnac's experiment without the introduction of a non-intertial cs.

\begin{figure}[htp]
\centering
\includegraphics[totalheight=0.4\textheight]{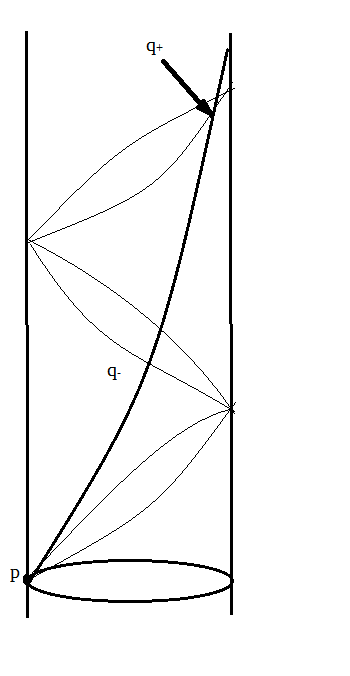}
\caption{Rotating ced}
\end{figure}

Let us call $\de_\pm (t)$ the maps representing the two opposite photons, for fixed $r$ and $z$, and $\g (t)$ the rotating ced; all are parametrized by means of $\phi$'s time coordinate:

\[ \g (t) = (r \cos\w t, r\sin\w t, z, t), \; \de_\pm (t) = (r\cos (ct/r), r\sin (\pm ct/r), z, t). \]

Let the departure event of the two photons be $p_0 \equiv (r,0,z,t)$. The two photons meet the ced when

\[ \pm \frac{ct}{r} \equiv \w t \; (\mbox{mod.}\, 2\pi), \]

\n
and in order to find the times of the {\sl first} meetings after departure, we must impose simply:

\[ \left\{\ba{rcl} \dss\frac{ct_+}{r} - 2\pi &=& \w t_+ \\ [8pt] -\dss\frac{ct_-}{r} + 2\pi &=& \w t_- \ea\right., \]

\n
that is:

\[ t_\pm = \frac{2\pi}{c/r\mp \w}. \]

\n
However, the time measured by the ced is related to the inertial time $t$ through the Proper Time Principle, and if we assume $\tau (0) = 0$ we obtain:

\[ \tau (t) = \frac{1}{c}\int_0^t |\dg (t)| dt = \sqrt{1-\frac{\w^2 r^2}{c^2}}t, \]

\n
and therefore \(\tau (t_\pm) = T_\pm\), as in the previous subsection. Thus the proper time difference is easily computed:

\be \De \tau = \tau (t_+) - \tau (t_-) = \frac{4\pi r^2 \w}{c^2}\frac{1}{\dss\sqrt{1-\frac{\w^2 r^2}{c^2}}}, \ee{sag_rel_pt}

\n
and we recover \rf{sag_rel} by putting $\bt = \w r/c$ and $A = \pi r^2$. 

\section{Instantaneous vs average speeds of light signals}

The computations in the previous section did not involve the average speeds of the photons with respect to the rotating platform. In order  to compute the average speeds of the photons, the rotating observer should measure the corresponding covered distances. However, the observer field 

\[ u (t,r,\te, z) =  \frac{(-\w r \sin (\w t +\te), \w r \cos (\w t +\te), 0, 1)}{\sqrt{1-\w^2 r^2/c^2}}, \]

\n
{\sl for which the instantaneous speeds of any photon is always $c$}, has no fixed 3-space. Strange as it may sound in classical terms, it can be said that in a rigorous sense the platform's rim {\sl does not exist as a spatially extended entity} (as opposed to a space-time entity) {\sl according to the rotating observer}, while it is a perfectly legitimate spatial entity (i.e. it occupies, under parallelism, the same region in parallel simultaneity spaces) {\sl according to the stationary observer}. This is just another one of the `relativities' in SR.

This difficulty arose early in the history of relativity, as Einstein strove to determine the spatial geometry of a rotating disc (cf. \cite{mm05a}, pp. 100-2). Einstein surmised that for the rotating observer the geometry of the disc is non-Euclidean, since the ratio of the `length of the rim' to the radius is bigger than $2\pi$. If we call (as above in \S 2) $\ell = 2\pi r$ the length of the rim according to the inertial observer, and $L$ the `length of the rim' according to the rotating observer, then we have, taking into account length contraction at the infinitesimal level and integrating: 

\be \ell = \sqrt{1-\w^2 r^2/c^2} L . \ee{rotd}

\n
This approach, though it gives a quantitatively correct result, misleadingly suggests that the platform lies in a single rest-space also for the rotating observer. 

A more logically satisfactory way of arriving to this formula for the rotating observer is to {\sl define} such a length as the arithmetical mean of $c \tau (t_\pm)$:

\[ L := \frac{1}{2} (c\tau (t_+) + c\tau (t_-)) = \frac{2\pi r}{\sqrt{1- \w^2 r^2/c^2}}, \]  

\n
in agreement with \rf{rotd}. By this procedure, one can make sense of the `length of the rim' also according to the rotating observer {\sl without having to postulate that for such an observer there be a `rim'}. We can also say that the `average speeds' of the two photons between departure and first meeting with the ced are for the rotating observer:

\be \tl{c}_\pm = \frac{L}{\tau (t_\pm)} = \frac{c}{1\pm \w r/c} = \frac{c}{1\pm\bt},  \ee{avel}

\n
one being subluminal and the other one superluminal, even though the instantaneous speeds are always in both cases $c$. This is paradoxical from a classical point of view, but the paradox dissolves once one realizes that in SR the platform, according to the rotating observer, cannot be construed as a fixed spatial entity. Misunderstanding of this point results in paralogisms such as the following (\cite{S10}, italics added):

\begin{quote}\small
Thus we see that, according to [SR] and respecting the isotropy of space, in every point of the rim the velocity of light relative to the disc is $c$ both clockwise and counterclockwise, indipendently of disk rotation. {\sl Therefore} two light pulses moving in opposite directions need the same time to complete the tour and the Sagnac effect goes to zero, contrary to empirical evidence.
\end{quote}\normalsize 

\n
The stated consequence would be correct only with a ced at rest in an inertial cs. But with respect to the rotating ced it is invalid, as `returning to the ced after emission' corresponds to two inequivalent worldlines, as Fig. 2 clearly illustrates. 

Notice that if we wish to express \rf{sag_rel_pt} in terms of $L$ we get simply:

\be \De \tau = \frac{2 L}{c}\bt, \ee{sag_rell}

\n
as in the first equality in \rf{sag_rel}.

A final remark is that from the simple fact that $\De \tau \neq 0$, the rotating observer can establish its own non-inertiality, with no information whatsoever gained from spatial coordinates (and of course no communication with the inertial observer). So in SR the Sagnac experiment can be correctly interpreted as analogous to Foucault's pendulum, as already stressed by Langevin \cite{lan21}.

\section{Nonstandard synchronizations}

Let us turn our attention to the issue of nonstandard synchronizations, which is crucial to Selleri's case for absolute simultaneity. We shall see that even this argument is marred by the illicit assumption described in the previous section.

If we synchronize from the stationary cs $\phi$ a clock on the rim travelling with velocity $\vy = \w r \uy$, where $\uy$ is a unit tangential vector at $\ry_0$ at the time $t_0$, and if we assume only the two-way speed isotropy of light, we can build an instantaneous cs $\phi_\vy$ of the form (cf. \cite{mm16}, \S 6):

\[   \phi_\vy :   x'= K\La (\vy) x +b, \; b = (\ry_0, t_0)\]

\n
where

\[ \La (\vy) :=  \left(\ba{cc} J & -\al \vy \\ -\frac{\al}{c^2 }\vy^T & \al \ea\right), \; J = I_3 + (\al -1)\uy\uy^T , \; K = \left(\ba{cc} I_3 & \0 \\ \ky^T & 1\ea\right), \]

\n 
with $\al = (1-\bt^2)^{-1/2}$ and  $|\ky |<1/c$ (compare with \rf{nons}). In the (3+1) formalism this amounts to: 

\be\left\{\ba{rcl} \ry' &=&  J\ry - \al t\vy +\ry_0 \\  [6pt] t' &=&  \al (1-\ky\cdot\vy)t + (J\ky - \al\dss\frac{\vy}{c^2})\cdot\ry +  t_0,\ea\right. \ee{con} 

\n
As a special case, if we take $\uy = \ey_1$, $\ky \propto \ey_1$, $\vy\propto \ey_1$, then  

\be\left\{\ba{rcl} x'^1 &=&  \al (x^1 - vt) +x_0^1 \\  [4pt] x'^2 &=& x^2 + x_0^2 \\ [4pt] x'^3 &=& x^3 + x_0^3 \\ [4pt] t' &=&  \al ((1-k_1 v) t +\dss \frac{x^1 }{c^2} (c^2 k_1 - v)) +  t_0,\ea\right. \ee{con_spe} 

Clearly the $\phi_\vy$'s time coordinate will be the usual one if and only if $\ky = \0$ (standard synchrony), while the synchrony established by $\phi_\vy$ will coincide with that of $\phi$ (`absolute' synchrony)  if and only if:

\[  \ky = J^{-1} \al\frac{\vy}{c^2}= \frac{1}{c^2} \vy.\]

The instantaneous speeds of light signals leaving $\ry_0$ in opposite tangential directions $\pm \uy$, according to a co-moving rotating observer at $\ry_0$, are obtained by substituting in \rf{con}  the formula $\ry (t) = \ry_0 \pm (t-t_0)c\uy$. Taking into account that:

\[ (J\ky - \al\frac{\vy}{c^2})\cdot\uy = \al (\ky\cdot \uy - \frac{\bt}{c})\]

\n
one obtains:

\be c (\pm \uy) = \frac{c}{1\pm c\ky\cdot \uy} \ee{two}

\n
which satisfies, as it should be, the two-way light isotropy condition: 
 
\be \frac{1}{c_+} + \frac{1}{c_-} = \frac{2}{c}. \ee{har}

In the special case considered above we have, for the instantaneous speeds of the two tangential light signals:

\be c(\pm\uy) = \left\{\ba{rcl} c\;\;\;\; & & \mbox{with standard synchrony} \\ [6pt] \dss\frac{c}{1\pm\bt} & & \mbox{with `absolute' synchrony}\ea\right. .
\ee{sell}

The second equality in \rf{sell} is exactly the formula \rf{avel} for the {\sl average} speeds of light according to the rotating observer as computed in \S 4. So one can state the following proposition, which contains what is right in Selleri's main argument:

\n
{\bf Proposition} {\sl The only choice of instantaneous synchronizations at each point of the rim of the rotating platform which guarantees that for the rotating observer the instantaneous speeds of the circling light signals coincide, respectively, with the average speeds is the synchronization imported from the stationary observer}.   \hfill $\Box$

Selleri holds that this proposition forces us to adopt `absolute synchrony' on the platform, instead of standard synchrony. We have seen in the previous section why this inference is wrong: the average speeds of the light signals for the rotating observer are definitely {\sl not} the length of a spatial path divided by a time of percurrence. For the stationary observer, on the other hand, they are just that, except that the clockwise and the counterclockwise paths are different, and need different times to be covered.  However, the real problem, which will be discussed in \S 10, is: how `absolute' is the synchrony Selleri is talking about?

\section{The zero acceleration discontinuity argument}

Suppose now that in order to comply with Selleri's stricture we, the rotating observer, were to adopt not just $\phi$'s synchrony, but the very time coordinate of $\phi$. Then we would have, from \rf{min_rot}:

\be \frac{d\te}{dt} = \left\{ \ba{cc} c/r -\w \\ [8pt]  -(c/r +\w) \ea\right. , \ee{angs_abs}

\n
and by multiplication by $r$ we would get, in module:

\be  \hc_\pm = \left\{ \ba{cc} c(1 -\bt)  \\ [8pt]  c(1 +\bt) \ea\right. .  \ee{sel1} 

\n
Notice that this is different from the `absolute' synchrony equality in \rf{sell}

\be c(\pm \uy) = \frac{c}{1\pm\bt} \ee{sel2}

\n
although the ratios coincide:

\be \frac{\hc_+}{\hc_-} = \frac{1-\bt}{1+\bt} = \frac{c(\uy)}{c(-\uy)}, \ee{sel3}

\n
and this is enough for what follows. 

In \cite{S97c} Selleri advanced an argument against the relativity of synchrony, which can be formulated by starting from the first equality in \rf{sel3}:

\be  \frac{\hc_+}{\hc_-} = \frac{1-\bt}{1+\bt}.    \ee{sel4}

\n
Clearly this formula does not change if $r$ increases while, correspondingly, $\w$ decreases as $v/r$: and yet - Selleri objects - for very big $r$ and very small $\w$ the observer on the rim of the platform is very nearly an inertial observer, so the ratio $c_+/c_-$, according to standard SR, should tend to 1 and not to the value given in \rf{sel4}. This is Selleri's ``zero acceleration discontinuity''. 

In fact SR has a simple answer to this objection: there is no discontinuity to eliminate, since, by the first equality in \rf{sell}, the ratio is always 1 when standard synchrony is adopted. Thus the objection strictly depends on the argument, which we have shown to be fallacious, based on the Proposition. For the rotating observer to use $\phi$'s synchrony would lead to the paradoxical equation \rf{sel4}. Indeed, Selleri's argument can be considered as evidence for the consistency of the SR's interpretation. 

On the other hand, for $\w \ra 0$,  from \rf{sag_rel_pt} we get $\De \tau \ra 0$, as it should be.

\section{Linear Sagnac effect}

Something very similar to the Sagnac's experiment can be obtained by restricting rotation to arbitrarily small time intervals \cite{wzyl03, wzy04}. 

\begin{figure}[htp]
\centering
\includegraphics[totalheight=0.1\textheight]{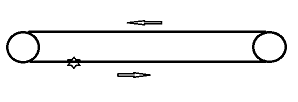}
\caption{Linear Sagnac effect}
\end{figure}

Let a conveyor belt carrying a ced be put in counter-clockwise motion, with an engine ensuring that the belt move with a speed $\pm v$ during the whole process except at the two ends. At the beginning of the experiment the ced emits light signals along the belt in opposite directions, and then, when they come back, registers their delay by the shift of interference fringes. 

\begin{figure}[htp]
\centering
\includegraphics[totalheight=0.4\textheight]{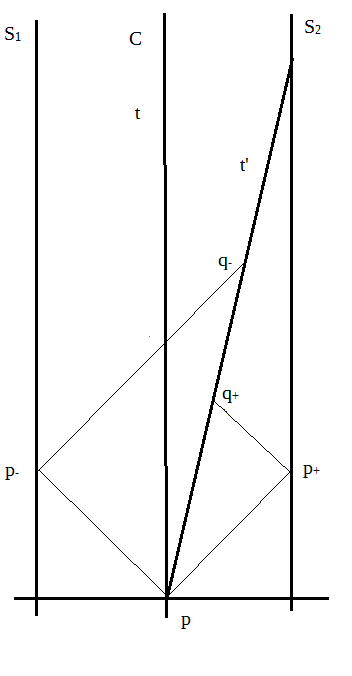}
\caption{Linear Sagnac effect in one spatial dimension}
\end{figure}

We can simplify the experiment by reducing the two direction changes to mirror reflections. 

An analogous and more `traditional' setting would be the following: let two mirrors $S_1$ and $S_2$ be placed on a rectilinear railway at a certain distance $2 L$; let a wagon move with a constant speed $V$ and let it emit two light signals towards $S_1$ and $S_2$, respectively, from the middle point of $S_1 S_2$; at the center of the wagon there is a ced which will measure the difference of the return times. Essentially, this {\sl wagon experiment} (as I shall call it) is the linear Sagnac experiment in a single space dimension.

In Fig. 4 we are assuming that $v$ is small enough for the backwards-moving signal to catch up with the ced before the ced reaches $S_2$, which means that 

\[ (0<) v\leq \frac{c}{3}. \] 

\n
Clearly in all foreseeable real experiments this inequality is likely to be largely satisfied. This also implies that for all the duration of a single (or even a few repetitions of the) experiment the observer is inertial.

Suppose the wagon's cs $\phi'$ is related to the railway's cs $\phi$ by the coordinate change (two spatial dimensions omitted with respect to \rf{con_spe}, and $x^1, x'^1$ re-written as $x, x'$) 

\be \left\{\ba{rcl} x' &=& \al (x- vt), \\ [4pt] t' &=& b (t+hx)\ea\right., \ee{ccc} 

\n
where 

\[ b: = \al (1-k_1 v), \; h: = \frac{k_1 - v/c^2}{1-k_1 v}. \] 

\n
In Fig. 4 the spatial axis for $\phi'$ is not indicated, as it might be just any spacelike line through $p$. By substituting $x = \pm ct$ in \rf{ccc} we obtain that the one-way speeds in $\phi'$ are:

\[ c_\pm = \frac{\al c}{b} \frac{1 \mp \bt}{1+hc}\]

\n
and by requiring that \rf{har} be satisfied, we obtain  \(b = \al (1-\bt^2)/(1+hv)\), which allows us to re-write  \rf{ccc} as: 

\be \left\{\ba{rcl} x'&=& \al (x- vt), \\ [4pt] t' &=& \al \dss\left(\frac{1-\bt^2}{1+hv}\right)(t+hx)\ea\right. . \ee{ccc1} 

Let $p_\pm ,q_\pm$, respectively, the reflection and return events of the positive and negative direction photons. Clearly

\[ p_\pm \equiv (\pm L, \frac{L}{c}), \]

\n
while a few computations show that:

\[ q_\pm \equiv (\frac{2L\bt}{1\pm\bt}, \frac{2L}{c}\frac{1}{1\pm\bt}) . \]

\n
By substituting these values into \rf{ccc1} we obtain:

\[ t'(q_\pm) = \frac{2L}{c}\al (1 \mp \bt), \]

\n
and therefore

\be \De t = \frac{4L\al}{c}\bt, \ee{sag_lin}

\n
which gives the same formula as \rf{sag_rel} (by substituting $2L = \ell$).

Let us remark that it was a priori clear that this formula could not depend on the choice of synchrony for the wagon, since $\De t$ is a proper time delay, and therefore an absolute effect. Accordingly, $\ep$ does not appear in the formula. In particular whether one accepts the conventional Lorentz transformation ($h= -v/c^2$, or $k_1 = 0$, or $\ep = 1/2$) or absolute synchrony ($h = 0$, or $k_1 = v/c^2$, or $\ep= (1+\bt)/2$), {\sl the predicted linear Sagnac effect remains the same}.

According to Selleri, the linear Sagnac effect proves that ``the physical difference between linear and curvilinear uniform motions vanishes'' \cite{S10}, and that therefore, with reference to \rf{avel}, also in inertial cs's not at rest in absolute space the speed of light should not be treated as isotropic. 

In fact, the formal analogy with the wagon experiment makes it intuitively clear why also in the standard Sagnac experiment the ced must register a delay in the return light signals: the antipodal point in the inertial frame, with respect to the position of the ced at the `double emission' event, can be thought of as `both mirrors in the wagon experiment, identified'. In the wagon experiment the mirrors are at rest in $\phi$, which is a different inertial system from the one of the wagon, so the fact that the instantaneous speed of light is always $c$ for the wagon is perfectly compatible with the round-trip spatial paths from and to the ced being different for the two light signals.

\section{Other arguments for absolute simultaneity}

In Selleri's papers on relativity one can find also other objections to relativistic simultaneity. Here are  a few.

A popular objection to the way the twin paradox is handled in SR is that the traveller twin is `almost always' in uniform motion and acceleration occurs only in a short, or even infinitesimally short, time interval, so how could a finite time difference be found? In fact this is as relevant as insisting that a broken line between two points in a Euclidean plane should not be longer than the segment joining those points since, after all, it deviates from a straightline only in the neighborhood of a few corners.

This specious argument has often crept again and again into discussions of the twin paradox, and in fact we find shadows of it in Selleri, for instance in the following statement (\cite{S10}, italics added), echoing and sharing a well-known claim by Herbert Dingle\footnote{``It should be obvious that if there is an absolute effect which is a function of velocity only, then the velocity must be absolute. No manipulation of formulae or devising of ingenious experiments can alter that simple fact.''  \cite{din57}}

\begin{quote}\small
Relativism [in handling the twin paradox]  does not apply and must be considered obsolete. Velocity ({\sl and nothing else}) is seen to be responsible for the differential retardation effect. Of course it must be an absolute velocity!
\end{quote}\normalsize

\n
Here the ``nothing else'' clause is meant to rule out any role for acceleration. Now, to say that `acceleration' is the cause of the differential ageing is an acceptable outline in the classical setting of the twin paradox (Fig.5). 

\begin{figure}[htp]
\centering
\includegraphics[totalheight=0.2\textheight]{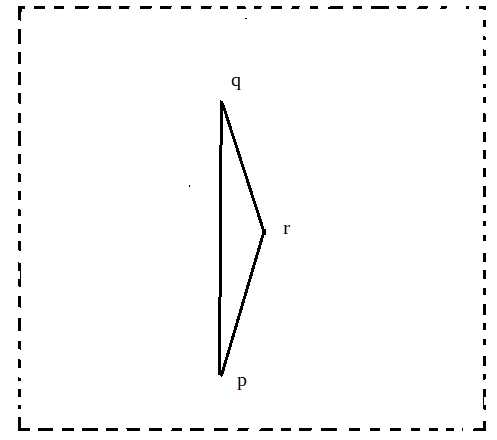}
\caption{Twins - classical setting}
\end{figure}

\n
But of course differential ageing would occur also in a generalized version of the paradox in which {\sl both} twins leave for sufficiently different accelerated travels before their final reunion (Fig. 6). 

\begin{figure}[htp]
\centering
\includegraphics[totalheight=0.2\textheight]{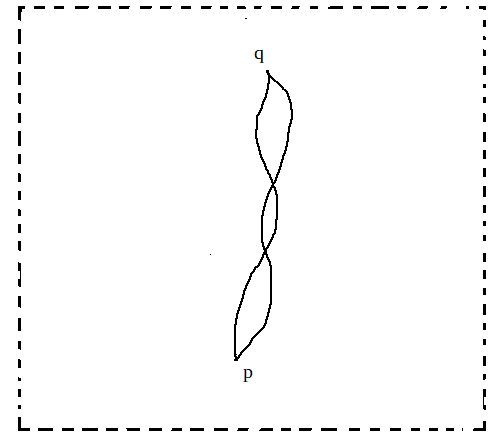}
\caption{Twins - generalized setting}
\end{figure}

\n
A slightly subtler case is when the twins move uniformly (but with different speeds) to a different place of the same Minkowskian cs - of course they must eventually decelerate to rest (Fig.7). 

\begin{figure}[htp]
\centering
\includegraphics[totalheight=0.2\textheight]{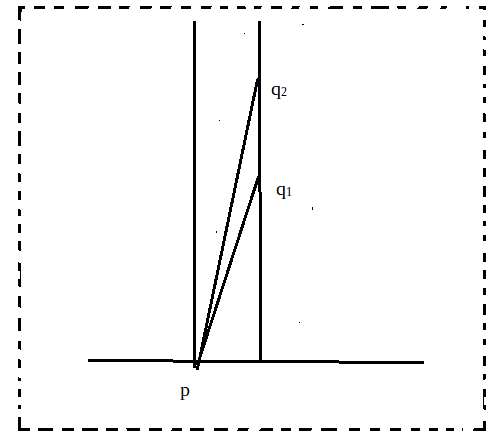}
\caption{Twins uniformly moving with different speeds}
\end{figure}

\n
In all three cases, clearly, reunion cannot occur unless some acceleration occurs. However, ultimately, the `cause' of the differential ageing is the Minkowski structure of space-time in SR, with the Proper Time Principle giving it a specific and absolute physical meaning in the case of timelike worldlines. It is this principle that bridges the gap between special and general relativity.  So ``relativism'' as in the first premise of the so-called ``Dingle's syllogism''\footnote{``According to the postulate of relativity, if two bodies (for example, two identical clocks) separate and reunite, there is no observable phenomenon that will show in an absolute sense that one rather than the other has moved.'' (\cite{din57}, p. 1242). To make justice to Dingle and numberless other critics of relativity, it must be conceded that some of Einstein's own formulations of `relativity' have been misleading (cf. \cite{mm05a}, pp. 117-9).} is indeed ``obsolete'', but all the same there is no need to postulate an absolute notion of rest.  

Another objection to SR was inspired by Popper \cite{pop} (see also \cite{pop1}; in fact the argument has a much longer history): special relativity ``leads one to accept a hyper-deterministic universe in which the whole future is completely pre-established in the minutest details and in which all sensations of individual freedom (even those limited to very simple events) are pure illusions'' \cite{S10}.  This is the {\sl Block Universe objection}. 

While a fuller account of it is out of place here, a short but sufficient reply can be given: whether the data on a spacelike slice are enough to determine all chronologically posterior events does not depend just on time order, but on the causal laws (which, particularly after quantum mechanics, are no more held in honour so much as they once were). As is well known, the most famous description of a hyper-deterministic universe had not to wait for the rise of SR, but had been provided by Laplace a century and a half earlier, within a perfectly Newtonian, if oversimplified (cf. \cite{true}, ch. 2), worldview \cite{lap}. 

A third objection was suggested to Selleri \cite{S02c, S10} by John Bell's lecture on how SR should be taught \cite{bell76}. Suppose that two spaceships are initially at rest in, and synchronized with, a Minkowskian cs $\phi$, and that they undergo identical accelerations until they are at rest in a different Minkowskian cs $\phi'$ (in a (1+1)-spacetime diagram their worldlines $\G_1$ and $\G_2$ can be seen as obtained one from a $x^1$-translation of the other, Fig. 8). It is clear that, when they `land' on $\phi'$,  their proper times coincide neither with $\phi$'s (that is, $t_0$) nor with $\phi'$'s time coordinate, and yet they are still synchronous with respect to $\phi$, not $\phi'$ (in fact $t'_2 < t'_1$).

\begin{figure}[htp]
\centering
\includegraphics[totalheight=0.3\textheight]{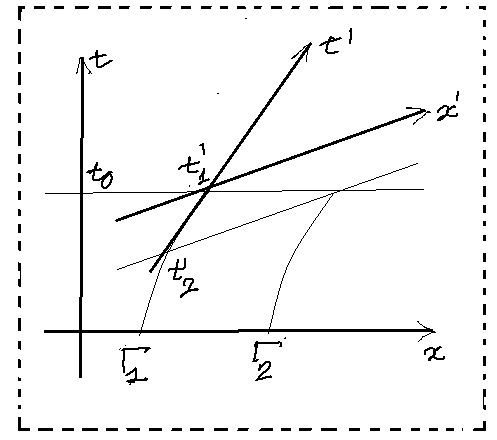}
\caption{Spaceships landing on a different cs}
\end{figure}

So far so good. It is hard to see, however, why this fact should be taken as contradicting SR. It is perfectly possible to adopt in a given notion of rest \cite{mm01} a synchrony imported from an arbitrary Minkowskian cs, rather than $\phi'$'s synchrony - i.e. the standard synchrony obtained by light signalling and $\ep \equiv 1/2$. The double-spaceship example shows how this might be done by a physical procedure, but in any case the Lorentz transformation from $\phi'$ to $\phi$ would enable an observer in $\phi'$ to compute $\phi$'s time coordinate without much trouble. There might be good practical reasons for the spaceship travellers to keep their clocks as $\phi$-synchronous instead of $\phi'$-synchronous, nonetheless they would be able to discover in a number of ways, by internal means, that this is indeed the case, since they would come across a number of metrical anomalies'' \cite{mm01,mm18} - contrary to Selleri's claim that ``Everything is regular, instead, if they [= two ``homozygous twins'', one for each spaceship] do not operate any asymmetrical modification of the time shown by their clocks'' \cite{S10}. 

So Selleri's objections fails to deliver the intended conclusion that the double spaceship system {\sl should be} synchronized as $\phi$.    

\section{``Weak relativity''}

Once we have seen that there is no disease in the SR's handling of rotational motion, we must show why Selleri's therapy is ineffective. The most obvious point is that if every rotating platform should adopt the time coordinate of a local Minkowski cs, this would still not imply absolute synchrony - just as absolute space is not identified by the Foucault's pendulum effect (only the corresponding Galilean class is). We should still have to make a further choice among the infinitely-many Minkowski synchronies - in other words we would still need to find a rationale in order to select the right one. And there is no doubt that SR with a physically relevant absolute synchrony is a {\sl different theory} from the usual SR: conventionalism is not a bus that can be taken and then be left at a convenient stop. 

I had the opportunity to discuss personally with Selleri his first steps towards a reform of special relativity, and also edited a book containing a survey he drew at the end of the 1990's  of his research on it \cite{S99a}. However, my only printed comment on Selleri's work appeared in 2001 (\cite{mm01}, referring to \cite{S96a, S99a}), in the following passage (pp. 802-3, italics added):

\begin{quote}\small
Thus it is true that if we establish absolute simultaneity, then faster-than-light processes are allowed, but these will only be those faster-than-light processes which do not produce causality reversal in the privileged frame itself. [...] For this reason any argument purporting to show the need to establish absolute simultaneity must be conceived in such a fashion as to make it clear that it only holds for the time coordinate of {\sl some} Minkowski frames; if, on the other hand, the argument can be freely applied to {\sl any} Minkowski frame, then it does prove too much. [...] [I]n fact, it is hard to see how a privileged Minkowski frame could be selected on physical grounds without reintroducing some concept of a material aether. It is my opinion that it would greatly serve the cause of clarity to openly recognize that a revision of special relativity is envisaged, and not a mere reformulation. 
\end{quote}\normalsize

\n
In fairness to Selleri, a change had already occurred in his work before I published these critical remarks: he had stopped holding that different synchronizations were merely ``equivalent'' forms of the same theory, and stressed instead, somewhat ambiguously, that they were ``alternative'' theories (as in the very title of \cite{S99a}). 

However, Selleri never clarified the crux of the matter: what is exactly the privileged synchrony which is selected by a correct (in his sense) interpretation of Sagnac's experiment? Ten years later, in \cite{S09}, he wrote, somewhat defensively, what amounts to his reply to my objection (italics added):\footnote{Both from indirect evidence, and because \cite{mm01} appeared in one of Selleri's favourite journals, I have no doubts that he was aware of my critical remark, although he never seem to have referred to \cite{mm01} explicitly.}

\begin{quote}\small
[...] the best theory of the physics of space and time seems to be the one based on absolute simultaneity.

Nevertheless {\sl we must admit that our results may seem somewhat contradictory}. On the one hand they point to a theory of space and time in which such conceptions as absolute velocity, privileged frame and absolute simultaneity have a central role, while, on the other hand, {\sl relativism comes back in the arbitrariness of the choice of the ``privileged'' inertial reference frame}. In spite of the fact that our results are mixed, we must stress that both sides of the contradiction (namely, [$k_1 = V/c^2$] and weak relativity) are absolutely correct if the assumptions made in the first section are correct. 
\end{quote}\normalsize

\n
Unfortunately, there is no escape here: a privileged inertial frame, however weakly characterized, cannot be {\sl arbitrarily} chosen, {\sl and yet Selleri's arguments, if valid, would lead exactly to this conclusion}.

In \cite{S08} Selleri made a distinction between SR, renamed  ``strong relativity'', and ``weak relativity, stating merely the impossibility to measure the absolute velocity of the Earth''. I suppose he meant that, in general, no absolute constant velocity could be measured, not just the Earth's. This is Lorentz's (and partly also Poincar\'e's) version of the theory of relativity, more commonly called ``neo-Lorentzian relativity'' (cf. \cite{S04c} and \cite{mm18}, \S 9). However, he insisted that only absolute simultaneity ``gives a satisfactory explanation of the Sagnac effect'', and ``a reasonable description of aberration'' and of the twin paradox. In the end, as we have seen, Selleri went so far as to write that ``Special relativity [that is, in the standard version] is self-contradictory'' \cite{S12}, and that a return to absolute simultaneity was a necessary amendment to recover consistency - although a rather precarious consistency, as admitted in the previous quotation. 

\section{Concluding remarks}

The history of attempts to show SR internally inconsistent is long and it is common to see it lightly disparaged, if not outright ridiculed. It should not, since it has been precious in deepening our understanding of the ways SR deviates from classical intuitions - no matter if it also included obvious blunders.\footnote{A notorious one focuses on finite $c$ behaving as an ``infinitely great velocity'', a statement by Einstein (\cite{ei}, p. 48) often misconstrued; for an early and effective refutation see \cite{re59}, pp. 6-7.} In any case, one should not only point out that a contradiction exists, but also evaluate its damaging consequences in the actual usage of the theory. To put it bluntly, a proof that a hammer cannot be used as a screw-driver does not make it useless for other purposes. Similarly, finding a contradiction in a physical theory is evidence that the theory is of limited value and must be corrected, not that it has no value and must be extirpated.\footnote{This view, incidentally, was supported even by Dingle, for instance in his posthumously published \cite{din80}.} This approach to foundational studies may encourage fair assessment of critical arguments, and help to prevent implicit and explicit forms of censorship in the editorial and academic practice.  

In this paper I have shown that the basic arguments advanced by Selleri to the effect that the relativity of simultaneity is untenable on theoretical and/or logical grounds are faulty. In particular, Sagnac's experiment can be dealt with in SR with no more than the subtleties one should expect. In general, rotation requires that one adopts the admittedly counterintuitive approach that for an observer on the rim of a rotating circular platform there is no natural synchronization along the rim, and in this sense there is not even a `rim' as a spatial entity. This is on a par with such previously shocking, but by now familiar, statements as that there is not one `length' of a ruler, there is not one `duration' of a physical process, and there is not one `mass' of a particle. If there is a lesson that everyone concerned should have learned after a century of debate on relativity is that trying to import Newtonian notions and imagery into SR is sure to wreak conceptual havoc. Of course one may try to revive Newtonian concepts (this has been done frequently in cosmology), and in a pluralistic view of scientific theories such attemps should be evaluated each on its merits.

Very different issues are whether a theory with absolute synchrony is empirically viable, and, if so, whether we should prefer it to SR. One is not bound to prove relativity self-inconsistent to be entitled to hold the view that its use in physics should be restricted. From this point of view, it is curious that Selleri never accepted that quantum mechanics, as commonly interpreted, needs absolute synchrony, which might have been used to make a stronger case for his main thesis \cite{mm18}.        

Let me end on a more positive note. While being critical of Selleri's main claims as regards SR, I think that Selleri's research on the foundations of relativity (and the organization of, or participation in, meetings and conferences in this area) has been instrumental in renewing the interest of several physicists in a  field that risked to be left as a proper subject for philosophical hair-splitting and corresponding inflated claims of relevance (cf. \cite{mm12}), with little or no relevance to working scientists. In other terms, in his last part of his career Selleri came to play the role of previous physics gadflies such as Herbert E. Ives \cite{t_h} and Dingle \cite{ding}.\footnote{Useful historical accounts of these scientists' critical stance towards relativity are \cite{lal13} and \cite{hc93}.} In my opinion the emergence of established scientists who are not afraid to espouse unpopular viewpoints in their disciplines is vital to keep an healthy atmosphere in the scientific community, and specific criticism of some of their arguments does not detract from this important contribution.    

\vs\vs
\n
{\bf Acknowledgment} In 1998-99 I supervised the Italian equivalent of a master dissertation on Selleri's work on the foundations of SR by Laura Febbraro, a brilliant student of mine and later my wife, reaching similarly negative conclusions on Selleri's program.

\renewcommand\refname{1. Selleri}

\renewcommand\refname{2. Sagnac and others}

\end{document}